\documentclass[twocolumn,aps,prb]{revtex4}
\usepackage{graphics,graphicx}
\usepackage{dcolumn}
\usepackage{bm}

\newcommand{\Ref}[1]{Ref.~\onlinecite{#1}}

\def\eb{\begin{equation}}   
\def\ee{\end{equation}}     
\def\ea#1{\begin{eqnarray} #1 \end{eqnarray}}   

\def\shro{Schr\"odinger}

\def\H{\hat{H}}
\def\HZ{\hat{H_0}}

\def\ex{\hat{x}}

\def\pe{\hat{p}}

\def\ra{\rightarrow}

\def\of#1{\left(#1\right)}

\def\eq#1{Eq.~(\ref{#1})}
\def\eqs#1#2{Eqs.~(\ref{#1}) and (\ref{#2})}

\def\sof#1{\left[ {#1} \right]}

\def\H{\hat{H}}

\def\Ppm{\Psi_\pm}
\def\ord#1{{\cal O}(#1)}

\begin{document}

\title{Reconciling Semiclassical and Bohmian Mechanics: \\
I. Stationary states}

\author{Bill Poirier}
\affiliation{Department of Chemistry and Biochemistry, and
         Department of Physics, \\
          Texas Tech University, Box 41061,
         Lubbock, Texas 79409-1061}
\email{Bill.Poirier@ttu.edu}

\begin{abstract}

The semiclassical method is characterized by finite forces and
smooth, well-behaved trajectories, but also by multivalued
representational functions that are ill-behaved at caustics.
In contrast, quantum trajectory methods---based
on Bohmian mechanics (quantum hydrodynamics)---are characterized
by divergent forces and erratic trajectories near nodes, but also
well-behaved, single-valued representational functions. In this
paper, we unify these two approaches into a single method that
captures the best features of both, and in addition, satisfies
the correspondence principle. Stationary eigenstates in one
degree of freedom are the primary focus, but more general
applications are also anticipated.

\end{abstract}

\maketitle                 


\section{INTRODUCTION}

\label{intro}

Theoretical and computational physical chemists have long
sought reliable and accurate means of performing quantum
dynamics calculations for molecular systems, as quantum effects
such as tunneling and interference often play an important role
in such systems. If ``exact'' methods are required---i.e.,
numerical techniques for which the error bars may (in principle)
be reduced arbitrarily---the traditional approach has been to
represent the system Hamiltonian using a finite, direct-product
basis set. However, this approach suffers from
the drawback that the scaling of computational effort is
necessarily exponential with system
dimensionality.\cite{bowman86,bacic89,bramley93}

Recently, a number of promising new methods have emerged that
may spell an end to exponential scaling---or at the very least,
drastically reduce the exponent. The latter category includes various
basis set optimization
methods,\cite{poirier99qcII,poirier00gssI,yu02a,yu02b,wangx03b}
which have nearly doubled the number of degrees of freedom (DOFs)
that may be tackled on present-day computers, from about 6 to 10
DOFs. Very recently, the first basis-set method to defeat exponential
scaling (at least in principle) was
introduced.\cite{poirier03weylI,poirier04weylII,poirier04weylIII}
This method, which uses wavelets in conjunction with a phase space
truncation scheme, has been applied to model problems up to 15 DOFs,
and is easily extendible to higher dimensionalities---although
some technical issues vis-a-vis applicability to real molecular
systems must still be resolved.

A completely different approach to the exact quantum dynamics
problem may be found in time-dependent trajectory methods.
Although trajectory methods are extremely common in
molecular dynamics applications, they are almost always
classical, quasiclassical, or
semiclassical\cite{madelung26,vanvleck28,keller60,maslov,arnold,gutzwiller,littlejohn92,brack}---i.e.,
not exact, in the sense described above.
However, it is possible to perform exact quantum dynamical propagation
using trajectory-based methods. These so-called ``quantum trajectory
methods''\cite{lopreore99,mayor99,wyatt99,wyatt01b,wyatt} (QTMs)
are based on the hydrodynamical picture of quantum mechanics,
developed over half a century ago by
Bohm\cite{bohm52a,bohm52b} and Takabayasi,\cite{takabayasi54}
using even earlier ideas of Madelung\cite{madelung26}
and van Vleck.\cite{vanvleck28}

Trajectory methods of all kinds are appealing, because they offer
an intuitive, classical-like understanding of the underlying dynamics.
QTMs are especially appealing, however---not only because they ultimately
yield exact results, but also because they offer a pedagogical
understanding of quantum effects such as tunneling.\cite{lopreore99,wyatt}
Curiously, QTMs thus far have not fared so well as semiclassical
methods, vis-a-vis their treatment of another fundamental
quantum effect---interference. This issue is discussed in more
detail below, as it is of central concern for the present paper.
An in-depth comparison of interference phenomena is provided in an
intriguing article by Zhao and Makri.\cite{zhao03} Garashchuk and
Rassolov\cite{garashchuk02,garashchuk03} discuss the interesting
connection between QTM and semiclassical
propagators,\cite{maslov,gutzwiller,littlejohn92} in the
context of Herman-Kluk initial value
representations.\cite{herman84,kay94,miller01}

Perhaps the greatest attraction of QTMs, however, has been the promise
that they may be able to defeat exponential scaling. In any event, QTMs have
undergone tremendous development since their introduction in
1999---most notably within the last year or two. Much of the
early development centered around accurate evaluation of spatial
derivatives of the wavefunction,\cite{lopreore99,mayor99,dey98}
but with the introduction of local least-squares fit adaptive and
unstructured grid
techniques,\cite{wyatt99,wyatt01b,wyatt} this difficulty is now
essentially resolved.  This has paved the way for a number of
interesting applications of QTMs, including barrier
transmission,\cite{lopreore99} non-adiabatic dynamics,\cite{burant00}
and mode relaxation.\cite{bittner02b} Several intriguing
phase space generalizations have also
emerged,\cite{takabayasi54,shalashilin00,burghardt01a,burghardt01b,donoso01}
of particular relevance for dissipative
systems.\cite{wyatt01,donoso02,bittner02a,hughes04}

On the other hand, QTMs still suffer from one major
drawback---they are numerically highly unstable in the
vicinity of nodes. This ``node problem'' manifests
in several different ways:\cite{wyatt01b,wyatt}
(1) infinite forces, giving rise to kinky, erratic trajectories;
(2) compression/inflation of trajectories near wavefunction local
extrema/nodes, leading to;
(3) insufficient sampling for accurate derivative evaluations.
In the best case, this can result in substantially more trajectories
and time steps than the corresponding classical calculation;
in the worst case, the calculation may fail altogether, beyond
a certain point in time.

For many molecular applications (though certainly not all),
the initial wavepacket is nodeless; however, it may develop
nodes at some later point in time. Moreover, from a practical
standpoint, nodes per se are not the only source of numerical
difficulty; in general, any large or rapid oscillations in the
wavefunction---termed ``quasinodes''\cite{wyatt}---can
be sufficient to cause the problems described above. Such
oscillations are very easily formed in molecular systems,
particularly during barrier reflection. Accordingly, several
numerical techniques are being developed to address the node problem,
including the Arbitrary Lagrangian-Eulerian frame
method,\cite{trahan03,kendrick03} and the artifical
viscosity method.\cite{kendrick03,pauler04}

In this paper, we take a different approach to the node problem,
based on a thorough understanding of the differences and similarities
between Bohmian and semiclassical mechanics.
Formally, these two theories share many similarities,
as was known from the earliest
days\cite{vanvleck28,morette52}---yet in practical terms,
semiclassical and quantum trajectories often behave very
differently. For instance, the former may cross in position
space, but not in phase space; the latter do exactly the
opposite. For the special case of stationary eigenstates in
1 DOF (the focus of the present paper),
semiclassical trajectories evolve in phase space along the
contours of the classical Hamiltonian, whereas quantum
trajectories {\em do not move at all}. For well-behaved
potentials, classical trajectories are always smooth and well-behaved,
but quantum trajectories may be kinky and erratic.

As noted by Zhao and Makri,\cite{zhao03}
nowhere are the differences between the two methods
more profound than in the treatment of nodes and
interference phenomena---which is not even {\em qualitatively}
similar. In the semiclassical treatment, the approximate wavefunction
is expressed in terms of simple functions that are generally as
smooth and well-behaved as the potential itself.
Oscillations and nodes are obtained when different lobes or
``sheets'' of the semiclassical wavefunction come to occupy the
same region of space---thus giving rise naturally to interference.
In contrast, the Bohmian representation of the wavefunction
is single-valued, and therefore incapable of self-interference.
Consequently, all of the  undesirable---even ``unphysical''---aspects
of quantum trajectories, as described in the previous paragraph,
are necessary in order to represent nodes and quasinodes explicitly.

From a dynamical perspective, the only difference between
semiclassical and quantum trajectories is the quantum potential,
$Q$, and it is a remarkable fact that $Q$ alone is responsible for
all of the very fundamental differences described above. On
the other hand, this situation is also cause for alarm, for $Q$
turns out to be the order $\hbar^2$ correction that is ignored in
semiclassical treatments---being regarded as ``insignificant''
in the large action limit, in accord with the correspondence
principle. From the previous discussion, it is clearly incorrect
to regard $Q$ as insignificant, which seems to place the
semiclassical approximation---and indeed, the correspondence
principle itself---in jeopardy. Moreover, there are certain
unappealing aspects of the semiclassical approach---multivaluedness,
caustics, phase discontinuities,
etc.\cite{vanvleck28,maslov,arnold,gutzwiller,brack}---that simply do
not arise in a Bohmian treatment. On the other hand, the
semiclassical approximation is known to be valid in the large
action limit---which together with the undesirable features
of the Bohmian approach as discussed in the previous paragraph,
seems to call into question the physical correctness of the latter.
This paradox has been a source of concern for some researchers,
notably Einstein.\cite{holland,floyd94,brown02}

The primary goal of the present paper is to reconcile the
semiclassical and Bohmian theories, in a manner that preserves
the best features of both, and also satisfies the correspondence
principle. At least within the context of stationary eigenstates
in 1 DOF, the above paradox turns out to be remarkably easy to
resolve. It can be shown that the disturbing dissimilarities
described above stem not from the theoretical methodologies themselves,
but from the simple fact that each method uses a different
ansatz to represent the wavefunction---thus giving rise to
a rather unfair comparison. In particular, since the semiclassical
functions are double-valued in the classically allowed region of
space, the stationary wavefunction is represented as the sum of two
terms---essentially a pair of ``traveling waves,'' headed in opposite
directions.  In contrast, the standard Bohmian approach uses
a single term to represent the wavefunction. Virtually all of the
disparities described above arise from this simple fact.

It is therefore natural to consider what would
happen if the Bohmian formalism were applied to a {\em two-term}
wavefunction---thus placing it on a proper
par with the semiclassical method. As will be shown in this paper,
this results in everything ``falling into place.'' In particular,
the quantum potential---far from being singular in the vicinity of
nodes---is well-behaved everywhere, and in fact, becomes vanishingly
small in the large action limit, exactly in accord with the
correspondence principle. The same can be said for the quantum
trajectories, which are no longer stationary, and approach the
corresponding semiclassical trajectories in the large action limit
(within the classically allowed region of space). This implies the somewhat
counterintuitive result that quantum trajectories must be
well-behaved when the number of nodes is {\em large}, for this
signifies the large-action limit.
In any event, the two-term Bohmian approach provides us with the
``best of both worlds,'' i.e. the well-behaved trajectories of
semiclassical mechanics, together with the singlevaluedness and lack
of caustics and phase discontinuities that characterize Bohmian
mechanics. More generally than for just the stationary states in
1 DOF considered here, it is anticipated that a multi-term Bohmian
implementation will go a long way towards alleviating the node problem.


\section{Background}

\label{background}

\subsection{Unipolar ansatz}

Let $\Psi(x)$ be any normalized wavefunction in the single
DOF, $x$. Being a complex function, $\Psi(x)$
can be decomposed into two real functions, $R_B(x)$ and $S_B(x)/\hbar$,
representing the amplitude and phase, respectively, as follows:
\eb
     \Psi(x) = R_B(x) e^{i S_B(x)/\hbar} \label{oneLMB}
\ee
Equation~(\ref{oneLMB}) is the celebrated Madelung-Bohm
ansatz,\cite{madelung26,bohm52a,bohm52b} which we term the
``unipolar ansatz,''\cite{wyatt} as it
consists of a single term only. The function $S_B(x)$ has units
and interpretation of action. If time evolution is considered,
then $S_B(x)$ plays the role of Hamilton's principle function
in classical mechanics,\cite{goldstein}
which therefore properly depends on $t$ as well as on $x$.
However, for stationary states in a time-independent context,
$S_B(x)$ is analogous to Hamilton's characteristic function,
$W(x)$, which is time-independent. Both interpretations will
be found to be important for the present approach.

The above decomposition is essentially unique if $R_B(x)$ is
nonnegative throughout; however, it leads to
discontinuities in $S_B(x)$, and cusps in $R_B(x)$, if $\Psi(x)$
has nodes. Despite these drawbacks, \eq{oneLMB} is the decomposition
generally utilized in standard Bohmian
treatments,\cite{wyatt,bohm52a,bohm52b} thus
motivating use of the ``$B$'' subscript. It will be shown
in this paper---evidently for the first time---that this convention in and
of itself gives rise to certain node-related numerical difficulties
that would otherwise not arise (Sec.~\ref{nodeissues}).

Accordingly, for the present work, we presume amplitude functions
that change sign when passing through nodes. To avoid confusion with
the standard Bohmian decomposition, we use unsubscripted quantities
to represent the present unipolar ansatz decomposition, i.e.
\eb
     \Psi(x) = R(x) e^{i S(x)/\hbar}, \label{oneLM}
\ee
where $R_B(x) = |R(x)|$, $S_B(x) = S(x) \bmod(\pi \hbar)$, and
the new decomposition functions, $R(x)$ and $S(x)$, are smooth
and continuous throughout the entire coordinate range; for the latter
reason, $S(x)$ is deemed a better analog for Hamilton's functions
than is $S_B(x)$. In any event, throughout this work, when discussing
the unipolar ansatz, we are referring to \eq{oneLM}, unless explicitly
stated otherwise. Although the function $S(x)$ is unique modulo
$2 \pi \hbar$, physically, it is well-defined only up to the addition
of an arbitrary constant, which introduces an arbitrary but
immaterial phase factor into $\Psi(x)$.

Equation~(\ref{oneLM}) is the starting point of both quantum trajectory
{\em and} semiclassical methods. Presuming a quantum Hamiltonian of the form
\eb
     \H = {\pe^2 \over 2m} + V(\ex), \label{Ham}
\ee
the general time evolution equations for the \eq{oneLM}
unipolar decomposition functions are
\ea{
\dot R & = & {-1 \over 2m} \of{2 R' S' + R S''}  \label{Rdotuni} \\
\dot S & = & - V  + {1 \over 2m}
   \left [ \hbar^2 \of{{R'' \over R}} - {S'}^2 \right ], \label{Sdotuni}}
where $R' = d\sof{R(x)}/dx$, $\dot R= d\sof{R(x)}/dt$, etc., and the
coordinate dependences have been suppressed to save space.
Equation~(\ref{Rdotuni}) is the continuity
equation, essentially stating that probability is conserved.
Note that this equation is independent of the
particular system potential, $V(x)$.

Equation~(\ref{Sdotuni}) is the quantum analog of the Hamilton-Jacobi
equation,\cite{goldstein} which does depend on the particular $V(x)$.
It is in the treatment of this equation that the semiclassical and
Bohmian theories part company. The former ignores the first term
in the square brackets,
giving rise to the standard classical Hamilton-Jacobi
equation. The latter regards the first term as the ``quantum potential,''
\eb
     Q(x) = - {\hbar^2 \over 2m} \of{R''\over R},
     \label{Qex}
\ee
which is added to the true potential, to form
the modified potential, $U(x)$. Apart from the substitution
$V(x) \ra U(x)$, the dynamical laws for the two approaches are
identical.

In particular, in both cases, the momentum $P(x)$ is related
to the action via
\eb
     P(x) = S'(x)
     \label{peeeq}
\ee
The set of points $\{x,P(x)\}$ constitute a one-dimensional subspace of
the 1 DOF phase space known as a ``Lagrangian manifold''
(LM).\cite{keller60,maslov,arnold,littlejohn92}
This terminology is generally used in a classical or semiclassical context
only; however, we find it convenient to apply it in the exact quantum
case as well.

\subsection{Stationary states, and the bipolar ansatz}

\label{stationarybipolar}

If $\Psi(x)$ is presumed to be a stationary state---i.e., an eigenstate
of the Hamiltonian, $\H$, with energy $E$---then $\dot R = 0$, and
$\dot S = -E$. The first result, together with \eq{Rdotuni},
is consistent with the well-known property that for standing waves,
the phase is constant over $x$. Without loss of generality, we may
take $\Psi(x)$ to be real, so that $S(x)=0$ and $R(x) = \Psi(x)$.
The second result, together with \eq{Sdotuni}, yields the
Quantum Stationary Hamilton-Jacobi Equation (QSHJE).
By rearranging \eq{Sdotuni}, and making use of \eqs{Qex}{peeeq},
we obtain:
\ea{
P^2(x) & = & 2m\sof{E - V(x) - Q(x)} \nonumber \\
       & = & 2m\sof{E - U(x)} \label{QSHJE} }
An important connection between the semiclassical and Bohmian
theories is suggested by \eq{QSHJE}---namely, {\em the semiclassical
approximation is accurate when the quantum potential is small}.

Let us now consider the semiclassical approximation proper---i.e.,
\eq{QSHJE} with no $Q(x)$ term. The resultant algebraic equation
has two solutions for $p(x)$, i.e.
$p(x) = \pm \sqrt{2 m [E-V(x)]}$---where lower case `$p$' is now used,
for reasons that will be explained shortly. These two solutions
correspond to the positive and negative momentum ``sheets'' of the
LM in phase space. The two sheets are joined together at the
classical turning points, $x_{\text{min}}$ and $x_{\text{max}}$,
to form a single LM in phase space, along the classical
Hamiltonian contour, $H(x,p) = p^2/2m + V(x) = E$. (Turning points
are the caustics for stationary states). Thus,
the function $p(x)$ is double-valued over the classically allowed
region, $x_{\text{min}} \le x \le x_{\text{max}}$, and zero-valued
everywhere else.

It is illuminating to compare the semiclassical situation, as
described above, with the Bohm prescription.
From the correspondence principle, one expects $Q(x)$ to be small
in the large action limit---i.e., the limit in which the excitation
numbers, $n$, become large. This in turn would imply, via \eq{QSHJE},
that the semiclassical and exact quantum LMs would resemble each
other in the large action limit. The actual situation is completely
different, however. Firstly, \eqs{oneLM}{peeeq} imply that
the quantum $P(x)$ is single-valued everywhere---rather than
zero- or double-valued, like the semiclassical $p(x)$.
Secondly, for stationary states, $S(x) = P(x) = 0$, implying
that the quantum LM is the real axis of the phase space plane,
which in no way resembles the semiclassical, Hamiltonian-contour
LM. Thirdly, quantum trajectories are stationary over
time, whereas semiclassical trajectories evolve around
their LMs (Sec.~\ref{trajectories}).

The origin of these seemingly profound qualitative differences is
deceptively simple: it is due to the fact that the semiclassical
approximation does not actually incorporate the unipolar ansatz
of \eq{oneLM}. Instead, a {\em bipolar} ansatz is used for the
total wavefunction, consisting of two terms rather than one:
\ea{
     \Psi(x) & = & e^{i \delta} r(x) e^{i s(x)/\hbar} +
               e^{-i \delta} r(x) e^{- i s(x)/\hbar} \nonumber \\
             & = & \Psi_+(x) + \Psi_-(x)
     \label{twoLM}}
In \eq{twoLM} above, the constant $\delta$ specifies the relative
phase between the two components, $\Ppm(x)$. It has been singled
out from $s(x)$ in order to clarify certain issues pertaining to
square-integrability (Sec.~\ref{quantization}). Although $\Psi(x)$ itself
is real, the $\Ppm(x)$ are both complex, and conjugate to each other.
The stationary ``standing wave'' is therefore obtained in practice
as the linear superposition of two ``traveling waves,''
moving in opposite directions.

The standard Bohmian prescription, being unipolar, completely misses
out on this elegant and useful aspect of the semiclassical approach,
which gives rise to a qualitatively very different kind of
amplitude/action decomposition.  On the other hand, if the
bipolar ansatz of \eq{twoLM} is incorporated into the Bohmian
theory, rather than \eq{oneLM}, then it is indeed possible to
reconcile these two approaches, in a manner consistent with the
correspondence principle, as will be shown in Sec.~\ref{decomposition}.
Note, however, that an important difference between semiclassical
and bipolar quantum LMs is already evident in \eq{twoLM}; namely
that the bipolar momentum function $p(x) = \pm s'(x)$ is double-valued
throughout the entire coordinate range. Thus, the two exact quantum
LM sheets never join, but extend into the classically forbidden regions
all the way to $x = \pm \infty$.

Throughout this paper, we use lower case to denote the bipolar ansatz
functions, so as to distinguish these from the unipolar ansatz functions,
for which upper case is used. For convenience, all bipolar functions are
hereafter defined to be single-valued everywhere, by referring to the
positive-momentum LM sheet only---e.g., $p(x) = s'(x)$.

\subsection{Node issues}

\label{nodeissues}

The increased flexibility of the bipolar ansatz is
extremely useful vis-a-vis the treatment of nodes, for it allows
for the direct representation of nodes as {\em interference fringes
arising naturally between the two traveling waves}. This
possibility is exploited to great effect in semiclassical methods,
which manage to contrive (approximate) bipolar amplitude functions
$r(x)$ that are completely {\em nodeless}---no matter how many
nodes are present in $\Psi(x)$ itself. Thus, apart from discontinuities
near turning points (associated with Maslov
indices\cite{keller60,maslov,gutzwiller,littlejohn92}),
the semiclassical $r(x)$ is smooth and positive, and the semiclassical
$s(x)$ is smooth and monotonically increasing. Moreover, these decomposition
functions tend to be very slowly varying, in relation to $\Psi(x)$ itself,
particularly when the latter has many nodes.

The above properties would of course also be beneficial for
exact QTMs---which from a practical standpoint,
is a primary reason why the bipolar ansatz ought to be considered
within a Bohmian context.
One difficulty is that the exact quantum bipolar decomposition of
\eq{twoLM} is {\em not unique}, in the sense that the QSHJE of
\eq{QSHJE} has a two-parameter family of
solutions.\cite{floyd94,brown02}
In particular, one trivial solution is
$r(x) = \Psi(x); s(x) = 0 = \delta$, which simply reproduces the
unipolar result. Clearly, this is not the solution that we want,
i.e., one that exhibits semiclassical correspondence in the
large action limit; obtaining the latter will be the focus of
Sec.~\ref{decomposition}.

For the unipolar ansatz, it has been stated that
nodes always give rise to infinities in $Q(x)$,
owing to the singular denominator in \eq{Qex}.
However, this is only strictly true if the standard Bohmian
ansatz of \eq{oneLMB} is used, for which the $R_B(x) \ge 0$
convention is employed. If instead, one adopts the \eq{oneLM}
convention, so that $R(x)$ smoothly changes sign as a node is
traversed, then \eq{Qex} need not always be singular at a node.

In particular, $Q(x)$ is {\em never} singular when $\Psi(x)$ is a
stationary eigenstate of $\H$, provided $V(x)$ is well-behaved
everywhere. This is because the time-independent \shro\ equation
guarantees that the nodes of $\Psi(x)$ also happen to be inflection
points. Using \eq{Qex} moreover, it can be shown that
$Q(x) = E - V(x)$.
Thus, for the stationary states considered in this paper,
even the unipolar ansatz is not singular, contrary to what previously
has been widely considered to be the case. Even
if the standard Bohmian ansatz is used, $Q_B(x)$ at a node
does not exhibit a singularity per se, but is rather
ill-defined, owing to the cusp in $R_B(x)$;
away from the nodal point, $Q_B(x) = Q(x)$.

In any event, we find it useful and convenient to distinguish between
two types of nodes, depending on whether $Q(x)$ is formally well-behaved
(``type one'' nodes) or singular (``type two'' nodes). From a numerical
perspective, even type one nodes will cause difficulties for standard
quantum trajectory calculations performed using the unipolar Bohmian
ansatz. This is because the slightest numerical error in the evaluation of
the \eq{Qex} ratio will result in instability near the nodes---even
though formally, $Q(x)$ does not diverge. In contrast, due to the smoothness
and lack of nodes of the $r(x)$ functions that arise in the bipolar
decomposition, numerical evaluation of the corresponding bipolar quantum
potentials, $q(x)$, causes no such instabilities for type one nodes.
More general nodal implications of the bipolar ansatz will be discussed
in greater detail in future publications.


\section{Bipolar decomposition for stationary states}

\label{decomposition}

\subsection{Semiclassical properties}

\label{semiprops}

In this section, we derive a unique bipolar decomposition
of the \eq{twoLM} form, for any given stationary wavepacket,
$\Psi(x)$, which satisfies semiclassical correspondence
in the large action limit. In general, the \eq{twoLM}
decomposition is nonunique. The semiclassical solution,
however, is essentially unique (Sec,~\ref{stationarybipolar}).
We will therefore use the latter as a guide,
for selecting the particular quantum bipolar
decomposition which most closely resembles the semiclassical
solution.

Note that certain assumptions have already entered
into the form of \eq{twoLM}, which is clearly more
constrained than a completely general bipolar decomposition
of $\Psi(x)$ into two arbitrary components. In particular,
we have presumed $\Psi(x)$ to be a superposition of equal and
opposite traveling waves---a natural assumption, completely
analogous to the semiclassical situation. Be that as it may,
there is still an enormous number of ways in which
\eq{twoLM} may be be realized for a given real
wavepacket $\Psi(x)$, and so the decomposition is still
far from being unique.

To help narrow the field, we first summarize some of the
additional properties of semiclassical eigenstates
in 1 DOF, which we will attempt to emulate in the exact
quantum decomposition:
\begin{enumerate}
\item{The LM itself {\em does not change} over time.}
\item{The classical probability distribution
{\em does not change} over time.}
\item{The classical flow for either LM sheet maintains
{\em invariant flux} over all $x$, with the flux value
for the two sheets being equal and opposite.}
\item{The area enclosed within the LM, i.e. the enclosed action, $J$,
is given by $J = 2 \pi \hbar (n+1/2)$, where
$n$ is the number of nodes.}
\item{For a normalized distribution, the absolute value of the
invariant flux equals the inverse of the period of the trajectory, i.e.
$F = \omega/2\pi$.}
\item{The median of the enclosed action, $x_0$, satisfies
$\int_{x_{\text{min}}}^{x_0} p(x) dx =
 \int_{x_0}^{x_{\text{max}}} p(x) dx$. }
\item{All trajectories move {\em along} the LM.}
\end{enumerate}
By ``translating'' these properties appropriately into the exact
quantum context, we will be able to define an essentially
unique bipolar decomposition of the \eq{twoLM} form.

\subsection{Basic properties: (1)--(4)}

\label{eigenstates}

\subsubsection{invariant flux}

\label{invariant}

Properties (1) and (2) above are the most fundamental,
and will be considered first. For a particular \eq{twoLM}
decomposition, the corresponding positive-momentum LM sheet
is given by $p(x) = s'(x)$. Property~(1) states that
$\dot p(x) = 0$, which in turn implies $\dot{s}(x) =\text{const}$.
Property (2) implies $\dot r(x) = 0$. Together, these properties
imply that the $\Ppm(x)$ {\em components} of $\Psi(x)$ must
be stationary eigenstates of $\H$ in their own right.
Equation~(\ref{twoLM}) then implies that the eigenvalues for
the two components must both be equal to $E$.

Since the $\Ppm(x)$ are stationary, the quantum mechanical
flux associated with each of these components, i.e.
\ea {
     j_\pm(x) & = & {\hbar \over 2 i m}
         \sof{\Ppm^*(x) {d\Ppm(x) \over dx} -
              {d\Ppm^*(x)\over dx} \Ppm(x)} \nonumber \\
     & & = \pm \sof{p(x) \over m} r^2(x), \label{fluxeq}}
is independent of $x$, with equal and opposite constant values,
$\pm F$. Note that \eq{fluxeq} demonstrates that the quantum
analog of property (3) is also satisfied---a
necessary consequence of properties (1) and (2). We call
this the ``invariant flux'' property.

For $F=0$, $\Psi_+(x) = \Psi_-(x) \propto \Psi(x)$,
which reproduces the unipolar ansatz. We shall therefore
hereafter restrict consideration to the $F>0$ case, for which
the invariant flux property, and \eq{fluxeq}, provide a
specification for $r$, in terms of $s'$, and the constant, $F$:
\eb
r(x) = \sqrt{\frac{m F}{s'(x)}} \label{RFeq}
\ee
Note that if $r(x)>0$ for all $x$, then \eq{RFeq} implies
that $p(x) > 0$ for all $x$---a desirable property
for the positive momentum solution, also satisfied by the
semiclassical solution. This would also imply
that $s(x)$ is monotonically increasing. Accordingly,
the $r(x)>0$ condition is adopted.

For $F>0$, the two $\Ppm(x)$ components are linearly independent.
This implies that at least one of the two must be non-$L^2$.
In fact, being complex conjugates of each other, {\em both}
solutions must be non-$L^2$. Moreover, it can be shown
that $\Ppm(x)$ diverges as
$x \ra \pm \infty$.\cite{milne30,floyd86}
This is due to the fact that $p(x) = s'(x) \ra 0$ as
$x\ra \pm \infty$ in order that the enclosed action be finite
(Sec.~\ref{quantization}); but this implies via \eq{RFeq} that
$r(x)$ diverges.

It is instructive to rederive the invariant flux property in
another manner. By the superposition principle, the time
evolution of $\Psi(x)$ can be obtained by propagating each
of the $\Ppm(x)$ components separately in time. Since
the two components are stationary eigenstates of $\H$
in their own right,  the time-evolving $\Ppm(x)$'s must
each independently satisfy $\dot r = 0$ and $\dot s = - E$.
The former, applied to a lower case version of
\eq{Rdotuni}, is equivalent to the spatial derivative
of \eq{fluxeq}.

\subsubsection{quantization}

\label{quantization}

We now address the exact quantum analog of property (4), the quantization
condition. In a proper semiclassical treatment, this half-integer
condition on the enclosed
action,\cite{bohr13,wilson15,sommerfeld16} $J = 2 \pi \hbar (n+ 1/2)$,
must be supplemented by the discontinuous jumps in phase that
occur as one traverses a turning point, from one LM sheet to another.
In a certain sense, these
jumps account for the fact that the WKB solutions do not incorporate
the portion of the true wavefunction that tunnels into the forbidden
region---which contribute an additional one half quanta of action, over
the course of one complete circuit around the
LM.\cite{brack}
When this discontinuous contribution is properly added to the usual
enclosed action contribution, one obtains an {\em integer} quantization
condition for the total action, $J_{\text{tot}} = 2 \pi \hbar (n+1)$,
even within a purely semiclassical context.

\begin{figure}
\includegraphics[scale=0.5]{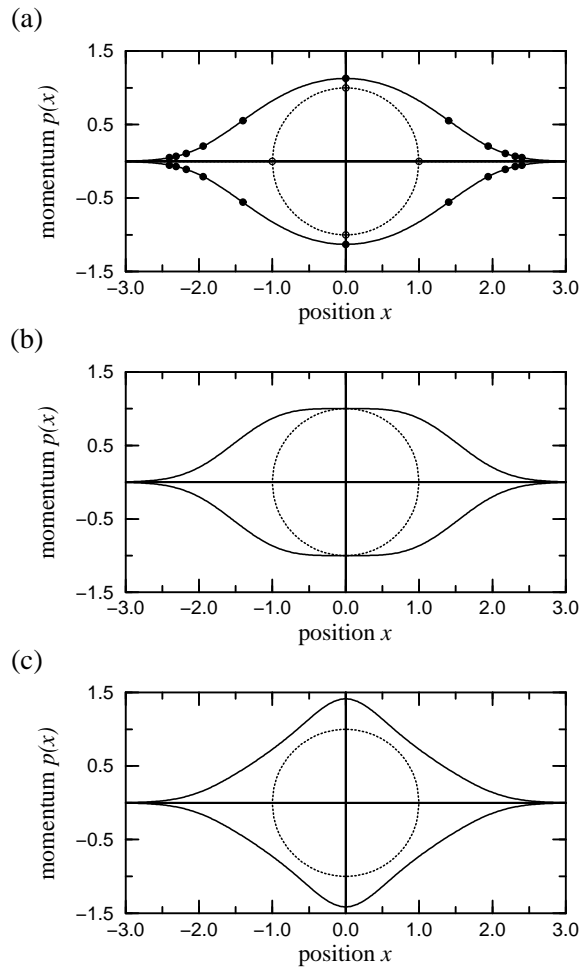}
        \caption{\label{HOgroundfig}
          Bipolar Lagrangian manifolds (LMs) for the harmonic oscillator
          ground state, $\Psi(x) = \exp(-x^2/2)/\pi^{1/4}$, for
          three different flux values, $F$ ($x_0=0$ in each case):
          (a) $F= 1/2\pi \approx 0.159$, the semiclassical value;
          (b) $F = 1/4\sqrt{\pi} \approx 0.141$;
          (c) $F = 1/2 \sqrt{2 \pi} \approx 0.199$.
          Solid curves indicate quantum LMs;
          dotted curves indicate semiclassical LMs. The former enclose an area
          $J = 2 \pi \hbar$; the latter $J = \pi \hbar$.
          In (a), the small open/filled circles represent semiclassical/quantum
          trajectory locations at times $t = k \pi/2$, for integer $k$.}
\end{figure}

In the quantum case, there is no distinction between classically
allowed and forbidden regions; one travels smoothly from one to the
other, over the entire position space. Since the two LM sheets
are symmetrically placed in phase space about the real
axis (Fig.~\ref{HOgroundfig}), the area enclosed between
them is clearly twice the change in the action function,
$\Delta s  = \int_{x_{\text{min}}}^{x_{\text{max}}} p(x) dx$,
as one travels from $x_{\text{min}}=-\infty$ to
$x_{\text{max}}=+\infty$. From the above
description, we expect this change in action to be $\pi \hbar (n+1)$,
where $n$ is the number of nodes. An integer quantization condition
is therefore expected to hold for the exact bipolar quantum
decomposition. This is indeed correct, as has been shown
previously.\cite{milne30,floyd86,poirier00qcI}

A sufficient, though certainly not necessary (see below),
condition for achieving integer quantization of the quantum action
is that $\Psi(x)$ be $L^2$, which requires that $\Psi(\pm\infty) = 0$.
For convenience, we adopt the convention that
$s(-\infty) = - \pi \hbar (n+1)/2$ in \eq{twoLM}. The condition
$\Psi(-\infty)=0$ then determines a value for $\delta$---i.e.,
$\delta = 0$ for $n$ even, and $\delta = \pi/2$ for $n$ odd.
This yields the following:
\eb
     \Psi(x) = \cases {2 r(x) \cos[s(x)/\hbar] & for $n$ even; \cr
                       2 r(x) \sin[s(x)/\hbar] & for $n$ odd. \cr }
     \label{twoLMsin}
\ee
The somewhat awkward distinction between the even and odd $n$ cases
is due to our choice of boundary condition for $s(x)$; the reasons
for this seemingly perverse choice will be made clear shortly.

Since $\Psi(+\infty)$ must also be zero if $\Psi(x)$ is $L^2$,
\eq{twoLMsin} implies that $s(+\infty) = \pi \hbar (n'+1)/2$,
where $n'$ is an integer. Since $r(x) > 0$ everywhere,
all nodes of $\Psi(x)$, for $n$ even/odd,  must occur at
$x$ values for which $s(x)/\hbar$ is an odd/even multiple of $\pi/2$.
This, in turn, requires $n'=n$, from which is obtained
\eb
     J =  2 \Delta s = 2 s(+\infty) - 2 s(-\infty) = 2 \pi \hbar (n+1),
     \label{quant}
\ee
i.e., the integer quantization condition. Note that $s(x)$ is
independent of the normalization of $\Psi(x)$. However, the
amplitude function $r(x)$ is not; thus, if $\Psi(x)$ is presumed
normalized to unity, then \eq{twoLMsin} defines the normalization
convention for $r(x)$, and for $\Ppm(x)$.

The specific value of the constant, $\delta$---as given above
for even/odd $n$---is not arbitrary, but is required in order to
ensure that the resultant $\Psi(x)$ be $L^2$. If a different value
were chosen, then a different, {\em non}-$L^2$ solution to
the \shro\ equation would be obtained. This situation is in stark
contrast to the unipolar case (for which the constant
simply changes the overall phase),
and is due to the fact that the bipolar $\delta$ represents a
{\em relative} phase shift. In any event, we can regard $\delta$
as a parameter that is used to specify a particular solution
of the \shro\ equation at the energy $E$. In fact, {\em all} such
real-valued solutions (apart from an immaterial scaling factor)
may be obtained by varying the value of $\delta$ in \eq{twoLM}.
This is due to the fact that the $\Ppm(x)$ are linearly independent.
In any event, an important consequence is that all \shro\
solutions are described by the exact same action function, $s(x)$,
which is unaffected by the value of $\delta$. Among other things,
this implies that the integer quantization rule applies to all
{\em non}-$L^2$ solutions, {\em as well as} to the $L^2$ solution.

\subsection{QSHJE properties: (5) and (6)}

\label{QSHJEprops}

\subsubsection{introduction}

\label{QSHJEintro}

In this section, we continue the approach introduced at
the end of Sec.~\ref{invariant}, where the time evolution
equations are applied to the two $\Ppm(x)$ components
separately. The $\dot r = 0$ equation was seen to yield
the same invariant flux relation, for each $\Ppm(x)$
component---i.e., \eq{RFeq}. The $\dot s = -E$ equation,
as applied separately to the two components, also yields
identical results. One of the important ramifications
of this is that the \eq{twoLM} ansatz is preserved over
time, at least for stationary eigenstates.

In any event, since $\Ppm(x)$ are solutions to the \shro\
equation, the $\dot s = E$ equation must result in a lower-case
version of the QSHJE [\eq{QSHJE}], for which
the bipolar quantum potential, $q(x)$, is defined via
lower-case \eq{Qex}. We shall rewrite this QSHJE
by expressing $q(x)$ directly in terms of $p(x)$, using
lower-case \eqs{Qex}{peeeq}, and \eq{RFeq}:\cite{messiah}
\eb
p^2 = 2m(E - V) - \hbar^2 \left [ {1\over 2} \of{{p''\over p}}
- {3\over 4}\of{{p'\over p}}^2 \right ]  \label{p2eq}
\ee
Whereas the semiclassical approximation [obtained by ignoring
the $q(x)$ term in brackets] has a unique $p(x)>0$ solution,
\eq{p2eq} is a second-order differential equation in $p(x)$,
with a two-parameter {\em family} of different solutions
to choose from.

Note that \eq{p2eq} applies to {\em all} \shro\ equation solutions,
i.e. the $L^2$ and the non-$L^2$ solutions, both. Since we are
interested only in the former, and since there is a one-parameter
family of \shro\ solutions in all, one might expect that the
specification of the $L^2$ solution would determine the value of one
of the two parameters. This is not correct however, as demonstrated
earlier (Sec.~\ref{quantization}). Thus, even for the $L^2$ solution
alone, bipolar decomposition gives rise to two variable parameters
via \eq{p2eq}.

\subsubsection{defining the two parameters}

What are the two parameters, and how should their
values be chosen?
To determine what the two parameters are, it is convenient to
combine \eqs{RFeq}{twoLMsin} together, to obtain a formula for
$\Psi(x)$ directly in terms of $s(x)$. The result is:
\eb
     \Psi(x) = \cases {\sqrt{{4 m F / s'(x)}} \cos[s(x)/\hbar]
                         & for $n$ even; \cr
                       \sqrt{{4 m F / s'(x)}} \sin[s(x)/\hbar]
                         & for $n$ odd. \cr}
     \label{Seq}
\ee
Equation~(\ref{Seq}) is a first-order differential equation for $s(x)$;
the general solution is easily found to be
\ea{
    \tan(s/\hbar) = {4 m F \over \hbar} \int_{x_0}^x {dx' \over \Psi^2(x')}
                        \qquad  & & \text{for $n$ even;}  \nonumber \\
    -\cot(s/\hbar) = {4 m F \over \hbar} \int_{x_0}^x {dx' \over \Psi^2(x')}
                         \qquad & & \text{for $n$ odd.} \label{tanS} }
For nodeless wavepackets ($n=0$), $\Psi(x)>0$ everywhere, and the
integrand of \eq{tanS} has no singularities.
When $n>0$, \eq{tanS} is still correct, but
requires careful branch selection, to ensure that the final
$s(x)$ curve is continuous throughout the coordinate range.
Note that $\Psi(x)$ is presumed to be the $L^2$ \shro\ solution.

Equation~(\ref{tanS}) provides an explicit recipe for obtaining
the \eq{twoLMsin} decomposition. The two parameters
can thus be taken as: (1) the flux parameter, $F$; (2) the
integration limit parameter, $x_0$. Note that $s(x_0) = 0$;
consequently, $x_0$ may also be interpreted as the median of the
action, as per Sec.~\ref{semiprops}. By varying the two parameters
$F$ and $x_0$ in \eq{tanS},
different bipolar decompositions may be achieved.
These correspond to different {\em affine transformations} of
each other, in the sense that varying $F$ is equivalent to
{\em rescaling} the right-hand-side (RHS) of \eq{tanS},
whereas varying $x_0$ is equivalent to {\em adding a constant}
to the RHS.

\subsubsection{choosing parameter values}

\label{paramchoice}

For a given $\Psi(x)$, the various bipolar LMs that can be
constructed via \eq{tanS} vary
significantly with respect to $F$ and $x_0$
(Sec.~\ref{HOresults}), and so a general
procedure for obtaining reasonable parameter values must be
provided. At present, the best approach seems to be to touch base
once again with the semiclassical properties---in particular,
property (5) for determining the appropriate value of $F$,
and property (6) for determining the appropriate value of $x_0$.

Semiclassically, the flux for a normalized distribution is
given by $F = \omega/2 \pi$, where $\omega$ is the classical angular
frequency for the appropriate semiclassical trajectory---i.e., the
(uniform) rate at which the angle variable of the action/angle
pair changes. The corresponding quantum trajectory is not that of a
bound state, and so it is not possible to assign an $\omega$ value
to it (Sec.~\ref{trajectories}). On the other hand, the \eq{twoLMsin}
normalization convention allows us to determine a unique flux
value for the quantum trajectory, which is all that is required.
By setting the quantum flux value equal to the semiclassical value,
it is anticipated that the resultant quantum LMs will closely resemble
the semiclassical LMs, as desired.

As for $x_0$, the median of the enclosed action: this can be
can be regarded as the exact middle of the wavepacket in a certain
sense; semiclassically, $x_0$ is the classically allowed configuration
that is furthest from both of the turning points, vis-a-vis the
action.  Consequently, we
expect the greatest agreement of semiclassical and quantum
LMs---i.e. the smallest $q(x)$ values---in the vicinity of the
semiclassical $x_0$. This can be achieved by allowing the quantum
$x_0$ to coincide with the semiclassical value---i.e., the
latter is chosen to be the location where $s=0$.

In the quantum bipolar decomposition scheme---even for fixed $F$---one
is otherwise free to place the action median, $x_0$, essentially anywhere
along the position axis. The ramifications are particularly
illuminating when $V(x)$ is even. For such potentials,
{\em only} the choice $x_0=0$ gives rise to quantum bipolar
decomposition functions that are even or odd in $x$, thereby respecting
the physical symmetry of the system, and of $\Psi(x)$ itself.
This choice for $x_0$ is also consistent with the median action
criterion. Presumably, it would be unphysical to consider any of the
asymmetrical decompositions; nevertheless, it is interesting to note
that one can generate {\em asymmetrical} bipolar decompositions that
give rise to the {\em symmetrical} $\Psi(x)$, simply by shifting
$x_0$ away from the origin. This has been verified via explicit
construction for the harmonic oscillator ground state.

\subsection{Quantum trajectories: property (7)}

\label{trajectories}

We now address the issue of the quantum trajectories themselves,
related to property (7). Semiclassically, over the course
of time, the bound state trajectories simply move around and around
the Hamiltonian contour LMs, which do not themselves change [property (1)].
In a conventional unipolar quantum treatment, the initial LM---specifying
the initial conditions for the ensemble of trajectories---is just the
real axis, i.e. the ``curve'' $P(x)=0$, since $\Psi(x)$ is real. The
quantum trajectories evolve under the unipolar modified potential,
$U(x) = V(x) + Q(x)$, which by \eq{Qex}, must be the constant function
$U(x) = E$ [even if there are nodes (Sec.~\ref{nodeissues})].
Consequently, $\dot P = -U'(x) = 0$, and so the unipolar quantum
trajectories {\em do not move at all} over time.

In contrast, the bipolar quantum trajectories are {\em not} stationary,
but move along the positive and negative momentum LM sheets. This
is true because $p(x) > 0$, and because the bipolar LMs themselves
do not change over time, thus verifying property (7).
Moreover, provided the bipolar quantum potential $q(x)$ is small in
the classically allowed region, then the bipolar quantum trajectories
must resemble the semiclassical trajectories within this region,
since the LMs are similar, and $\dot p = - u'(x) \approx -V'(x)$.
Of course, the bipolar quantum trajectories do {\em not} change their
direction at the classical turning points, moving between positive and
negative-momentum LM sheets, like classical trajectories.
Instead, all quantum trajectories on say, the positive momentum
LM sheet, continue to head to the right for all time.
Once these trajectories enter the classically forbidden region,
however, their speed decreases very suddenly.

It is worth discussing the very different role played by the
quantum potential in the unipolar ansatz, versus that of
the bipolar ansatz with the specific decomposition
suggested here (i.e., parameter choices of Sec.~\ref{paramchoice}).
In the unipolar case, $Q(x)$ serves to counteract the true
potential everywhere; thus, $Q(x)$ is not generally small.
In contrast, the bipolar quantum potential, $q(x)$,
can be regarded as the $\ord{\hbar^2}$ correction to the semiclassical
approximation---in the truest correspondence-principle sense of
lower-case \eq{QSHJE}.
The value of $q(x)$ is therefore small in the appropriate semiclassical
limits---i.e., in the classically allowed region far from turning points,
and in the limit of large action, when $n$ becomes large. Near
the turning points, $|q(x)|$ increases substantially, so as to
ensure that all trajectories keep moving past the classical turning
point without changing direction. This increase continues well into
the classically forbidden region, where curiously, $q(x)$ approaches
$Q(x)$---i.e. it effectively cancels out the true potential.
Consequently, the bipolar modified potential, $u(x) = V(x) + q(x)$,
resembles the true potential in the classically allowed region,
and the unipolar modified potential, $U(x) = E$, in the asymptotic
regions.

The above discussion hinges on the assumption that the semiclassical
and bipolar quantum LMs become arbitrarily close in the appropriate
semiclassical limits described above. We can justify this expectation
as follows. First, the semiclassical approximation is known to become
arbitrarily accurate in these limits; each of the two semiclassical
traveling wave components must therefore approach some particular
corresponding pair of exact quantum solutions, $\Ppm(x)$, arbitrarily
closely. The latter must therefore have the same characteristics,
vis-a-vis action, trajectories, and flux, as do the semiclassical
approximations, in the appropriate limits. Therefore, by choosing
the available parameters for the quantum solutions (i.e. $F$ and $x_0$)
so as to match the semiclassical approximations, the correspondence
principle must be satisfied.

\subsection{Stationary non-eigenstates}

\label{noneigen}

Although the primary interest of this paper is bound, stationary
eigenstates of the Hamiltonian $\H$ of \eq{Ham}, our ultimate
interest is wavepackets that evolve dynamically over time.
As a first step in this direction, we generalize the previous
discussion to include wavepackets that are only momentarily
``stationary.'' In other words, the initial
wavepacket $\Psi(x)$ is real, but otherwise arbitrary,
i.e., not presumed to be an eigenstate of $\H$. This results in
$\dot r = 0$, but only instantaneously, at the initial time.
We shall call such a wavepacket a ``stationary non-eigenstate.''

To what extent can the bipolar decomposition scheme be
applied to stationary non-eigenstate wavepackets?
The question is relevant, because it is only necessary to specify
the bipolar decomposition at a single point in time, in order to
propagate the two $\Ppm(x)$ components independently, over all
time.  Our approach shall be to regard $\Psi(x)$
as the eigenstate of some Hermitian, Hamiltonian-like
operator, $\HZ$, which without loss of generality, may be
taken to be of the form
\eb
     \HZ = {\pe^2 \over 2m} + V_0(\ex). \label{HZ}
\ee
If $\Psi(x)$ is known, it is a trivial matter to obtain $V_0(x)$ by
solving the \shro\ equation in reverse, i.e.
\eb
     \left[V_0(x)-E_0\right] = {\hbar^2\over 2m} \of{{\Psi''\over \Psi}}.
     \label{inverse}
\ee
The $\HZ$ so obtained can then be used to generate the appropriate
$F$ and $x_0$ values semiclassically.

As a completely general procedure, this approach has one unavoidable
flaw. If $\Psi(x)$ has type two nodes, then $V_0(x)$ will have
singularities at the nodes, which is undesirable.
In such cases, since the precise values of the $F$ and $x_0$ parameters
may not matter all that much in numerical practice, one should simply
choose ``reasonable'' values by comparison with known cases---e.g., for
$n=0$, one could use the parameters of a Gaussian with the same center
and standard deviation as $\Psi(x)$.  On the other hand, almost
all initial wavepackets used in chemical physics applications
correspond [via \eq{inverse}] to potentials $V_0(x)$ that are
well-behaved.


\section{Results: harmonic oscillator eigenstates}

\label{HOresults}

As a classic benchmark example, we now work out analytic solutions
for the harmonic oscillator (HO) eigenstates, i.e. $V(x) = k x^2/2$.
This example is particularly important, as the ground state
provides the proper decomposition for Gaussian wavepackets,
which are used very frequently in time-dependent studies. We shall
also find the excited harmonic oscillator states to be quite
useful, particular with respect to investigations regarding
nodes and interference. The normalized $n$'th harmonic oscillator
eigenstate is given by
\begin{eqnarray}
\Psi_n(x) & = & \left(2^n n!\right)^{1/2} \left(\frac{m \omega}{\hbar \pi}\right)^{1/4}\times\nonumber\\
     & & H_n \of{\sqrt{{m \omega \over \hbar}} x} e^{-m \omega x^2/2 \hbar},
\end{eqnarray}
where $H_n(\,)$ is the $n$'th Hermite polynomial, and
$\omega = \sqrt{k/m}$.

\subsection{Ground state}

\label{HOground}

We start with the ground state,
$\Psi_0(x) = (m \omega / \hbar \pi)^{1/4} e^{-m \omega x^2/2\hbar}$.
Application of \eq{tanS} yields
\begin{eqnarray}
s(x) & = & \hbar \arctan\left\{ F \left(\frac{2 \pi}{\omega}\right)\times\right.\nonumber\\
 & & \left.\left[\text{erfi}\left(\sqrt{\frac{m \omega}{\hbar}} x\right) - \text{erfi}\left(\sqrt{\frac{m \omega}{\hbar}} x_0\right)\right]\right\}
\end{eqnarray}
as the generic, $F$- and $x_0$-dependent solution. This gives rise
via \eq{RFeq} to
\begin{eqnarray}
r(x)  & = & \left(\frac{1}{2}\right) \left(\frac{m \omega}{\hbar \pi}\right)^{1/4} e^{-m \omega x^2/2 \hbar}\times\nonumber\\
 & & \left\{1 + F^2 \left(\frac{2 \pi}{\omega}\right)^2\times\right.\\
 & & \left.\left[\text{erfi}\left(\sqrt{\frac{m \omega}{\hbar}}x\right) - \text{erfi}\left(\sqrt{\frac{m \omega}{\hbar}}x_0\right)\right]^2\right\}^{1/2}.\nonumber
\end{eqnarray}

As per Sec.~\ref{paramchoice}, the appropriate value
of $F$ is clearly $F = \omega/2\pi$. The appropriate value of $x_0$,
whether from symmetry considerations, or the more general
median action criterion, is clearly $x_0=0$.
With these choices for the parameter values,
we obtain the simpler result
\ea{
s(x) & = & \hbar \arctan \left [ \text{erfi}\of{\sqrt{{m \omega \over \hbar}} x} \right ] , \\
r(x) & = & \of{1\over 2} \of{{m \omega \over \hbar \pi}}^{1/4} e^{-m \omega x^2/2 \hbar}\times\nonumber\\
& & \left [ 1 + \text{erfi}^2\of{\sqrt{{m \omega \over \hbar}} x} \right ] ^{1/2}. }
For convenience, we choose units such that
$\hbar = m = k = \omega = 1$. In these units, $F = 1/2\pi$, and $x_0=0$.
In these units, and for these parameter values, all of the relevant
bipolar decomposition functions are as follows:
\ea{
s(x) & = & \arctan\left[\text{erfi}\of{x} \right ] \nonumber \\
r(x) & = & \of{1\over 2}\of{{1\over\pi}}^{1/4} e^{-x^2/2}
\left [ 1 + \text{erfi}^2(x) \right ]^{1/2} \nonumber \\
p(x) & = & { 2 e^{x^2} \over \sqrt{\pi}
           \left [ 1 + \text{erfi}^2(x) \right ] } \nonumber \\
q(x) & = & {1 \over 2} - {x^2 \over 2} -
{ 2 e^{2 x^2} \over \pi \left [ 1 + \text{erfi}^2(x) \right ]^2 }
 \label{groundsum} }

All of the functions in \eq{groundsum} are smooth,
slowly varying, and monotonic in $|x|$. The LM is
an elegant ``eye-shaped'' curve [specified by the $p(x)$
equation above] that deviates smoothly, and positively, from the
circular semiclassical LM, with the point of closest approach being
$x_0=0$. All of these features are as predicted in
Sec.~\ref{decomposition}, and would not have been satisfied
if substantially different parameter values were used.
A plot of the semiclassical and bipolar quantum LMs is presented
in Fig.~\ref{HOgroundfig}, for $F=1/2 \pi$ and other values
(but all with $x_0=0$). Whereas some of these other plots have the
qualitatively correct behavior, it is very clear that the
$F= 1/2\pi$ curve is the smoothest, most ``correct''
choice---especially vis-a-vis comparison with the
corresponding semiclassical LM.

\begin{figure}
\includegraphics[scale=0.5]{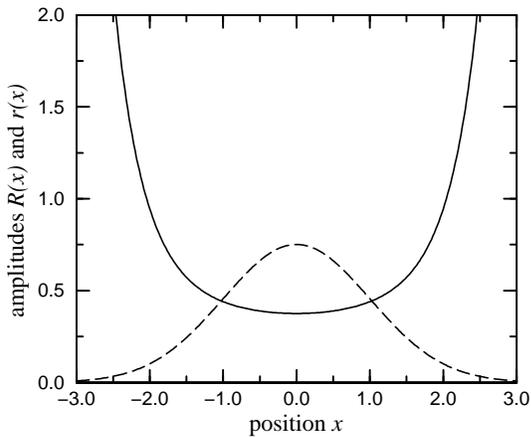}
        \caption{\label{HORfig}
          Amplitude functions for the harmonic oscillator ground state,
          $\Psi(x) = \exp(-x^2/2)/\pi^{1/4}$. Dashed curve: unipolar
          amplitude, $R(x) = \Psi(x)$. Solid curve: bipolar amplitude,
          $r(x)$, for semiclassical parameter values,
          $F=1/2\pi$ and $x_0=0$.}
\end{figure}

Figure~\ref{HORfig} is a comparison between the unipolar
and bipolar amplitude functions---i.e., $R(x)$ and
$r(x)$, respectively. As is clear from the figure, these two
types of amplitude behave completely differently. In particular,
whereas $R(x)$ decreases quickly as $x \ra \pm \infty$, $r(x)$
increases as one moves away from the origin, and actually diverges
in the $x \ra \pm \infty$ limits, as predicted in
Sec.~\ref{invariant}. Clearly, the $\Ppm(x)$ are non-$L^2$ solutions.

\begin{figure}
\includegraphics[scale=0.5]{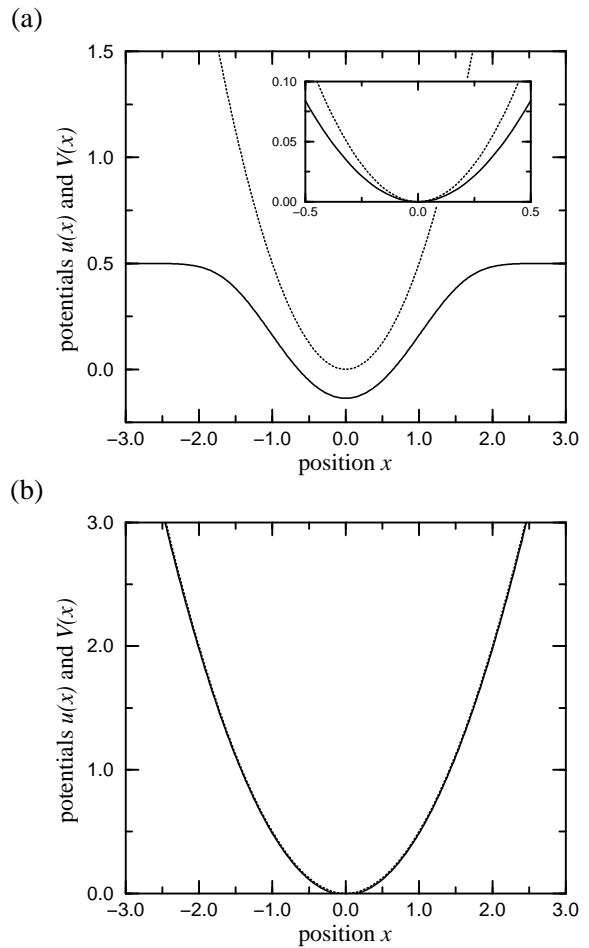}
        \caption{\label{HOpotfig}
          Bipolar modified potentials, $u(x)$, and true potential, $V(x)$,
          for two different harmonic oscillator eigenstates:
          (a) ground state, $n=0$; (b) tenth excited state, $n=10$.
          Solid curves indicate $u(x)$; dotted curves indicate $V(x)$.
          In (a), $u(x)$ approaches $E_{n=0}=1/2$ as $x \ra \pm \infty$,
          but resembles $V(x)$ in the classically allowed region, $|x|<1$
          [inset shows $u(x)-u(0)$ vs. $V(x)$]. In (b), $u(x)$ and $V(x)$ are
          virtually indistinguishable over the
          (classically allowed) coordinate range indicated.}
\end{figure}

In Fig.~\ref{HOpotfig}(a), we present a comparison of the actual
and bipolar modified potentials---i.e., $V(x)$, and
$u(x) = V(x) + q(x)$. The two potentials resemble each other in the
classically allowed region, away from the turning
points at $x=\pm 1$. As one approaches the turning points, the
difference $q(x)$ increases markedly. In the classically
forbidden region, $u(x)$ ceases to emulate the true potential,
and in the asymptotic limits, approaches the unipolar $U(x)$
constant value of $E=1/2$.
All of this is as predicted in Sec.~\ref{trajectories}.

We also performed trajectory calculations. In particular, a
single trajectory was propagated over a very long period of time,
using what Wyatt has called the ``analytical approach.''\cite{wyatt}
In this scheme, the modified force [i.e. $-u'(x)$] is computed
analytically, but the trajectory itself is propagated numerically.
We found first of all that this numerical propagation scheme
was extremely stable, as demonstrated by the fact that the numerical
trajectory did not deviate appreciably from the LM at any point in time.
Phase space values for the trajectory at various times are indicated
as small circles in Fig.~\ref{HOgroundfig}(a), from which it is also clear
that quantum trajectories correspond fairly well to the semiclassical
trajectories in the classically allowed region.

In the classically forbidden regions, trajectories
slow down very quickly, as predicted. This is evidenced by the
pile-up of trajectory points that ensues in these regions
[Fig.~\ref{HOgroundfig}(a)]. Formally, however, the trajectories do not
actually reach zero momentum until $x \ra \infty$. They are thus
analogous to classical trajectories for a system that has just enough
energy for dissociation. This fact is also reflected in the asymptotic
behavior of $u(x)$ as discussed above.

We now briefly address the issue of trajectory ``pile-up'' in the
classically forbidden regions, which is an important concern for
numerical calculations. Although the bipolar ansatz has the effect
of placing more trajectories in regions of space where the actual
probability is small, this situation is numerically agreeable for two
reasons: (1) more accuracy is needed in these regions, because
$\Psi(x)$ itself is effectively obtained as the difference between
two large numbers; (2) a simple ``recycling'' scheme can be
introduced to reduce the number of trajectories to a minimum. These
issues will be discussed in great detail in future
publications.

\subsection{Excited states}

\label{HOexcited}

The correspondence between the semiclassical and quantum LMs
for the harmonic oscillator ground state is only fairly good,
but one ought to recall that the action is minimal in this case.
A real test of the correspondence principle requires a detailed
investigation of the LM behavior in the large action limit. This in
turn, requires that the bipolar decomposition be performed for
the excited harmonic oscillator states. Using \eq{tanS},
with $\hbar = m = k = \omega = 1$, $F= 1/2\pi$, and $x_0=0$,
we have obtained analytical solutions for all $n$ up to $n=10$.

The general form of the bipolar action for the $n$th eigenstate
[denoted $s_n(x)$] is as follows:
\eb
s_n(x) =
\cases{ \arctan\left[{e^{x^2} f_n(x) \over \sqrt{\pi} g_n(x)}
    + \text{erfi}\of{x} \right ]  & for $n$ even; \cr
        -\text{arccot}\left[{e^{x^2} f_n(x) \over \sqrt{\pi} g_n(x)}
    + \text{erfi}\of{x} \right ]
                         & for $n$ odd. \cr}
\label{simpleex}
\ee
In \eq{simpleex} above, $f_n(x)$ is an $(n-1)$th-order odd/even
polynomial, and $g_n(x)$ is an $n$th-order even/odd polynomial,
for $n$ even/odd. Explicit coefficient values for $f_n(x)$ and
$g_n(x)$ are listed in Tables~\ref{ftab} and~\ref{gtab},
respectively (coefficients for larger $n$ can be provided on request).

\begin{table*}
\caption{\label{ftab}Coefficients, $a_j$, for polynomials,
$f_n(x) = \sum_{j=0}^{n-1} a_j x^j$, for all $0<n\le 10$.
These polynomials appear in the numerator of the analytic formula
for the bipolar action, $s_n(x)$ [\eq{simpleex}].}
\begin{ruledtabular}
\begin{tabular}{ccccccccccc}
Order & \multicolumn{10}{c}{Coefficients}
\\ \cline{2-11}  $n$ &
$a_0$ & $a_1$ & $a_2$ & $a_3$ & $a_4$ & $a_5$ & $a_6$ & $a_7$ & $a_8$ & $a_9$
\\ \hline
1  &  1 \\
2  && 2 \\
3  &  2 && -2 \\
4  && 10 && -4 \\
5  &  8 && -18 && 4 \\
6  && 66 && -56 && 8 \\
7  &  48 && -174 && 80 && -8 \\
8  && 558 && -740 && 216 && -16 \\
9  &  384 && -1950 && 1380 && -280 && 16 \\
10 && 5790 && -10560 && 4704 && -704 && 32 \\
\end{tabular}
\end{ruledtabular}
\end{table*}

\begin{table*}
\caption{\label{gtab}Coefficients, $b_j$, for polynomials,
$g_n(x) = \sum_{j=0}^n b_j x^j$, for all $0<n\le 10$.
These polynomials appear in the denominator of the analytic formula
for the bipolar action, $s_n(x)$ [\eq{simpleex}].}
\begin{ruledtabular}
\begin{tabular}{cccccccccccc}
Order & \multicolumn{11}{c}{Coefficients}
\\ \cline{2-12}  $n$ &
$b_0$ & $b_1$ & $b_2$ & $b_3$ & $b_4$ & $b_5$ & $b_6$ &
$b_7$ & $b_8$ & $b_9$ & $b_{10}$
\\ \hline
1  && -1  \\
2  & 1  && -2  \\
3  && -3  && 2 \\
4  & 3  && -12  && 4  \\
5  && -15  && 20 && -4  \\
6  & 15 && -90  && 60  && -8  \\
7  && -105  && 210 && -84 && 8  \\
8  & 105 && -840 && 840 && -224 && 16 \\
9  && -945 && 2520 && -1512 && 288 && -16 \\
10 & 945 && -9450 && 12600 && -5040 && 720 && -32 \\
\end{tabular}
\end{ruledtabular}
\end{table*}

The bipolar solutions for the excited states behave exactly as
predicted. In particular $r_n(x)>0$ everywhere, and both
$r_n(x)$ and $s_n(x)$ are as smooth, slowly varying, and
monotonic as for the ground state. In fact, all of the bipolar
functions qualitatively resemble those for $n=0$, except on a larger
scale---as is also true of the semiclassical functions. This is in
sharp contrast to the behavior of the wavefunctions, $\Psi_n(x)$,
themselves, which gain nodes and rapid oscillations as $n$ is
increased. One very encouraging aspect of the excited state
bipolar functions is that nodal features are not evident anywhere.
However, this requires choosing the correct branch of the \eq{tanS}
solution at every point in position space, such that the resultant
$s_n(x)$ curve is not discontinuous across a node. This issue is
discussed in more detail in Sec.~\ref{numericalissues}.

\begin{figure}
\includegraphics[scale=0.5]{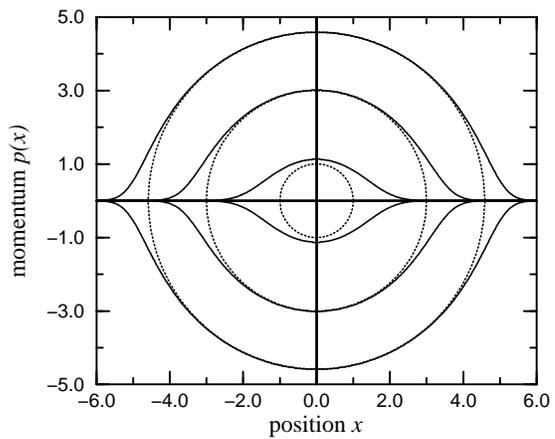}
        \caption{\label{HOexcitedfig}
          Bipolar Lagrangian manifolds (LMs) for three different harmonic
          oscillator eigenstates; moving concentrically outward from the
          origin, these are $n=0$, $n=4$, and $n=10$, respectively.
          Solid curves indicate quantum LMs;
          dotted curves indicate semiclassical LMs.
          The correspondence principle is clearly satisfied with
          increasing $n$. }
\end{figure}

In any case, the basic goal of the bipolar ansatz has been
achieved---i.e., to obtain a decomposition which, like the
corresponding semiclassical approximation, treats all nodes
as interference between a superposition of left and right
traveling waves, $\Ppm(x)$, which are themselves nodeless.
Moreover, beyond achieving just this basic goal, we find that the
{\em correspondence principle is satisfied in the large
$n$ limit}. This is exemplified in Fig.~\ref{HOexcitedfig},
wherein the semiclassical and bipolar quantum LMs are presented
for several harmonic oscillator states over the $n$ range considered.

From the figure, the quantum LMs are seen to enclose one-half
quanta of area more than the semiclassical LMs, which manifests
primarily in the forbidden regions near the turning points, as
expected. In a relative sense, this discrepancy
becomes decreasingly significant in the large $n$ limit.
Note that the quantum LMs completely enclose the semiclassical
LMs, which was not anticipated earlier, but is certainly
a desirable property. From lower-case \eq{QSHJE}, this
will only be satisfied if the bipolar quantum potential is
negative everywhere. This
has been observed for all examples considered thus far,
using the appropriate semiclassical values for $F$ and $x_0$;
however, most other parameter choices would not satisfy this
property.

Even more so than in the ground state case, the bipolar quantum
potential for $\Psi_n(x)$ is found to be very small throughout
most of the classically allowed region. The magnitude
of $q_n(x)$ decreases with increasing $n$,
so that whereas $q_0(0)= -.137$ for the
ground state, by $n=10$ we have $q_{10}(0) = -.012$. Of course, the
extent of the classical region is also larger with increasing $n$;
thus by $n=10$, we find $u_{10}(x)$ to be practically
indistinguishable from the $V(x)$ over the range
$|x| \le 3$. The situation, depicted in Fig.~\ref{HOpotfig}(b),
can be regarded as another manifestation of the correspondence principle.

The correspondence principle also has important ramifications for
trajectory calculations. In particular, not only are the bipolar
quantum trajectories for large $n$ smooth and well-behaved throughout,
but in the classically allowed region, they are virtually indistinguishable
from classical trajectories. This has once again been verified by
performing analytical trajectory calculations for $n=10$, which were
found to be just as numerically stable as for the ground state---despite
the fact that $\Psi_{10}(x)$ itself has {\em ten nodes}. This bodes
very well for obviating the node problem in general.

\subsection{Numerical issues}

\label{numericalissues}

Although the bipolar decomposition functions---once
obtained---exhibit no special behavior in the vicinity of
nodes, it turns out that nodes complicate the determination
of these functions somewhat, vis-a-vis implementation of
\eq{tanS}. To begin with, let us imagine that---as in the current
harmonic oscillator case---an analytical expression for the
\eq{tanS} integral is available. For the moment, we also
take $n$ to be even. Note that the left and right sides of
\eq{tanS} must be infinite at the nodes. Thus, whereas the
exact analytical expression can be used across the entire
coordinate range, a new branch is encountered each time a node
is traversed. The specific branch of interest is specified
by the condition of continuity for $s_n(x)$, and by
$s_n(x_0)=0$.

A superficial difficulty is encountered for the odd $n$
states, for which there is necessarily a node at $x_0$.
Strictly speaking, this implies that the \eq{tanS} integration
must be singular. To circumvent this difficulty, we express
the RHS of \eq{tanS} as an indefinite analytical
integral, plus an arbitrary constant, $B$. Note that since
$x_0$ must lie at a node for odd $n$, $x_0$ can not serve
as the second parameter, for singling out the particular
solution of interest for \eq{p2eq}. We can, however, use
$B$ for this purpose. In particular, if $V(x)$ is even,
then only one value of $B$ gives rise to the requisite
odd $s_n(x)$ solution. More generally, i.e. for arbitrary $V(x)$,
we can still apply the oddness criterion locally. In
other words, it is easy to show that $B$ should in general
be chosen such that $s_n(x_0-\epsilon) = - s_n(x_0+ \epsilon)$
in the limit of small $\epsilon$. This technique bears
a resemblance to the Cauchy principal value method.\cite{arfken}

The above discussion presumes that an analytical integral
is available for the RHS of \eq{tanS}. Generally
speaking, this will not be the case, and we must consider
how to apply the above procedures when the integrations are
performed numerically. Fortunately, this is straightforward.
The general procedure is to pick an arbitrary integration
limit, $x_i$,
lying in between each adjacent pair of nodes (where
$x= \pm \infty$ are treated as ``nodes'' in this context).
Equation~(\ref{tanS}) is then applied to
each interval separately, generating a smooth, numerically
integrated function over the entire $i$'th interval, that is
correct to within an additive constant, $B_i$. The individual
$B_i$ values are then obtained, using the constraint
$s_n(x_0)=0$, and applying the Cauchy-like condition described
above across each of the nodes separately.

\begin{figure}
\includegraphics[scale=0.5]{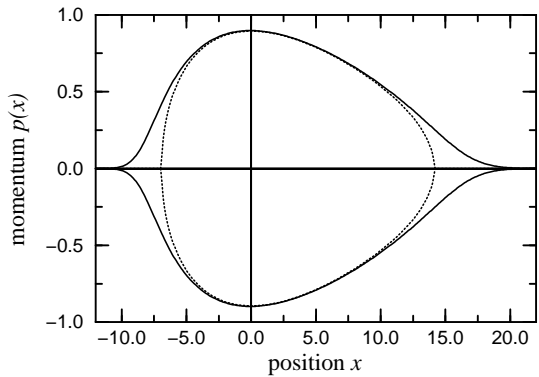}
        \caption{\label{morsefig}
          Bipolar Lagrangian manifolds (LMs) for the fourth excited
          eigenstate of the Morse oscillator system with twenty bound states
          total. Solid curves indicate quantum LMs; dotted curves indicate
          semiclassical LMs. Semiclassical values for $F$ and $x_0$ were used
          to specify the quantum solution, as per Sec.~\ref{numericalissues}.}
\end{figure}

The numerical procedure described above has been applied
succesfully to the Morse oscillator system, for which:
(1) $V(x)$ is not symmetrical, and; (2) the \eq{tanS}
integrations must be performed numerically. The results
will be presented in a future paper.
We mention the Morse investigation in this paper simply
to emphasize the fact the present method is in fact
applicable in a much broader context than the analytical
harmonic oscillator system considered here. Moreover,
all of the conclusions drawn for the harmonic oscillator,
regarding the correspondence principle and the like,
evidently apply to more general systems. This is demonstrated in
Fig.~\ref{morsefig}, which
depicts the semiclassical and bipolar quantum LMs for the
$n=4$ state of the Morse oscillator.

\section{Comparison with related methods}

\label{Floydcomparison}

The present work does not constitute the first application of
the bipolar ansatz in a Bohmian-like dynamical context. For
two decades or so, a very interesting bipolar approach has been
developed and advocated by
E.~R. Floyd.\cite{floyd94,brown02,floyd86,floyd82,floyd00,bouda01}
More recently, essentially the same technique was derived by
Faraggi and Matone (FM),\cite{brown02,faraggi99,faraggi00}
within a much broader context,
and using a very different physical picture.  It is somewhat
remarkable that these two approaches give rise to exactly the
same dynamical equations (an illuminating comparison is presented
in \Ref{wyatt} and in \Ref{brown02}). Perhaps even more
remarkable, however, is that the dynamical law used in
Floydian/FM trajectory propagation is {\em not} equivalent to that
of Bohmian mechanics.

In this section, we compare and contrast the methods of Floyd, FM,
and the present work. The various approaches were developed
independently, and so a brief discussion of the different
philosophies is presented, as well as the mathematical
similarities and differences.
The comparison is particularly apt for the present paper, in that
both the Floyd and FM theories are restricted to
{\em stationary states only}.

The starting point of the Floydian approach is the QSHJE in 1 DOF
(multidimensional systems may be considered, but only if there is
separation of variables\cite{brown02}). As per Sec.~\ref{background},
this is natural enough, for a Bohmian-like theory applied to stationary
states; however, there are some subtleties regarding the manner in
which energy is treated, that give rise to a non-Bohmian dynamical law.
The intriguing approach of FM begins not with the QSHJE, but with the
fundamental postulate that all systems are equivalent under coordinate
transformations.\cite{brown02,faraggi00} This is termed the ``equivalence
principle,'' in obvious analogy with general relativity, with which
the FM approach shares many parallels. Indeed, one of the goals of
FM is to reconcile quantum mechanics and general relativity, to which
end it is natural to focus on the action, which
plays a key role in both physical theories.

In any event, the FM approach relies on the existence of a particular
coordinate transformation that reduces the system to that of a
free particle.\cite{brown02,faraggi99} This approach implicitly relies
on the fact that the Wigner-Weyl correspondence is not
preserved under canonical/unitary
transformations\cite{poirier00qcI}---a feature that has
been exploited to great effect in improvements to semiclassical
theories, such as the Langer
modification,\cite{langer37,berry72} although
not always with full comprehension that this is what was
taking place.\cite{morehead95,poirier99lan}

Starting from the equivalence principle, FM {\em derive}
the QSHJE, and demonstrate that energy
quantization arises naturally from the condition that this
postulate be satisfied. They find, moreover, that a unipolar
ansatz is insufficient to achieve this, but rather, a bipolar ansatz
of the \eq{twoLM} form is required. In contrast, Floyd begins with the
QSHJE, and then presumes a bipolar ansatz for which each of the two
terms is a solution. Floyd's motivations are evidently similar
to the author's, in that he introduces the bipolar ansatz in
order to:\cite{floyd94,brown02}
(1) obtain a Bohm-like formalism that is well-behaved at nodes;
(2) obtain dynamical trajectories that are {\em not} stationary,
thus obviating Einstein's concern, and
allowing for the possibility of satisfying the correspondence
principle.

For purposes of this discussion, the approaches of Floyd and
of FM may from this point forward be regarded as identical.
Since both employ the QSHJE, as does the present approach, it is
clear that {\em all three utilize essentially the same bipolar
functions}, $s(x)$, $r(x)$, $p(x)$, and $q(x)$. However, there
are some important differences. Floyd uses a different
convention, for which both the normalization of $\Psi(x)$,
and the flux,
change simultaneously, so that his $r(x)$ is proportional to
ours. Neither of the other methods identifies the two parameters
associated with \eq{p2eq} in the way that we have done, and they
certainly do not provide a means of selecting preferred values for
these parameters, as per Sec.~\ref{paramchoice}.

Indeed, Floyd considers each of the two-parameter family of
solutions to \eq{p2eq}, which he terms ``microstates,''
to constitute an equally valid decomposition of $\Psi(x)$. He
further asserts that since the \shro\ equation per se provides
no means of distinguishing microstates, that the QSHJE must
be regarded as the more fundamental equation, in some sense.
Floyd also acknowledges the fundamental differences between
$q(x)$ and $Q(x)$, and the advantages of the former vis-a-vis
nodes, as discussed in Sec.~\ref{nodeissues} (although he
incorrectly claims that the latter is singular at the origin
for the first excited harmonic oscillator eigenstate\cite{floyd82}).
On the other hand, the lack of a criterion for selecting
a prefered microstate implies that the quantum trajectories
associated with his approach---termed ``Floydian trajectories''---will
not in general approach classical trajectories in the
correspondence principle limits.

Actually, there is another, more fundamental reason why Floydian
trajectories do not satisfy the correspondence principle. This is
because the underlying dynamical law governing their evolution is
radically different from that of the Bohmian quantum trajectories
utilized here. The trajectories are different, {\em even
though} the bipolar decompositions are identical---a very curious
situation that bears further analysis. In particular, the period of
Floydian trajectories is finite, necessitating a speed that approaches
infinity as $|x| \ra \infty$. In contrast, Bohmian trajectories
{\em rapidly slow down} in the forbidden regions, never reaching
the coordinate asymptotes, as discussed in Sec.~\ref{HOground}.
Whereas Floyd finds implications for hidden variables and the
Copenhagen interpretation, we adopt a decidedly
less philosophically ambitious perspective.

The Floydian dynamical law is obtained from the QSHJE
by applying a classical procedure,\cite{goldstein}
wherein trajectory evolution is related to the quantity
$\partial s / \partial E$ to yield
\eb
\dot x  = \of{1 - {\partial u \over \partial E}}^{-1}
{1 \over m} s'(x), \label{floyd}
\ee
where $u(x) = V(x) + q(x)$ is the bipolar modified potential.
Various interpretations may be provided for the
$(1 - \partial u / \partial E)$ factor in \eq{floyd}. FM adopt
a relativistic interpretation which lumps it
together with $m$ to form the ``effective quantum mass,''
whereas Brown (in essence) uses it to define an
``effective time.'' In the Bohmian approach, $u(x)$
is regarded as independent of $E$, thus giving rise to the
usual $p = m \dot x$ relation.  In the Floydian approach
however, $u(x)$ is considered to depend on $E$, giving rise
to the more complicated \eq{floyd} expression above,
for which the canonically conjugate momentum, $p(x)$, is not
necessarily equal to mechanical momentum, $m \dot x$.

How can the same $u(x)$ be regarded as energy-independent in one
theory and energy-dependent in another? It has to do with
different interpretations of the energy. One can make a distinction
between the quantum energy, $E_{\text{Q}}$, and the classical energy,
$E_{\text{C}}$. The quantum energy is that of the eigenstate
$\Psi$, which determines $q(x)$, $u(x)$, and
$h(x,p) = p^2/2m + u(x)$. The contours of $h(x,p)$ define the
classical energies $E_{\text{C}}$, with $E_{\text{Q}} = E_{\text{C}}$
corresponding to the bipolar quantum LM. The function $u(x)$ clearly
depends on $E_{\text {Q}}$, but does not depend on $E_{\text {C}}$.
The meaning of \eq{floyd} therefore depends on whether
$E = E_{\text {Q}}$ (Floydian approach), or
$E = E_{\text {C}}$ (Bohmian approach).

Floydian dynamics, therefore, represents
a different philosophical outlook than that of Bohmian dynamics.
Both approaches are correct, and ultimately yield mathematically
identical results. The main point we wish to make here is simply
that this difference does not appear to have anything to do with
the bipolar ansatz per se. Indeed, one could easily apply the
Floydian dynamical law to the {\em unipolar} ansatz---although
for stationary eigenstates, this would yield the same results
as Bohmian propagation.

We close this section with a few final comparisons between
the Floyd/FM and present approaches. First, we comment that
quantum energies are discrete for bound states, and so any
derivative with respect to $E_{\text{Q}}$ requires careful
consideration, as has been previously noted.\cite{brown02,floyd82}
This is relevant for Floydian trajectories, but is essentially
a non-issue for Bohmian trajectories. Second, although Floyd
has certainly considered the 1 DOF harmonic oscillator system, he
appears not to have derived analytic expressions for the relevant
bipolar functions, as we have done in Sec.~\ref{HOresults}.
We have, moreover, verified that Floyd's closed form expression
for the bipolar modified potential,\cite{floyd86}
\eb
     u(x) = E - 1/\sof{a \Psi_1(x)^2 + b \Psi_2(x)^2 +
     c \Psi_1(x) \Psi_2(x)}^2, \label{floydclosed}
\ee
is consistent with \eq{groundsum}, by substituting
$\Psi_1 = 2 r \cos(s)$, $\Psi_2 = 2 r \sin(s)$,
$a=b=\pi/\sqrt{2}$, $c=0$, and $E=1/2$ into \eq{floydclosed}.
Finally, we comment that the Floyd/FM approach, being
based on the QSHJE, does not appear to generalize to
the non-stationary-case, whereas the present approach
does---at least in certain situations.


\section{SUMMARY AND CONCLUSIONS}

\label{conclusion}

The \shro\ equation is linear, yet the equivalent
\eqs{Rdotuni}{Sdotuni}---obtained via substitution
of the \eq{oneLM} ansatz into the \shro\ equation---are not.
Quite apart from the philosophically intriguing issues which
this raises, a primary conclusion of the present work is that
this situation may also be exploited for practical purposes. In
particular, the superposition principle allows us to divide up
the initial wavepacket $\Psi(x)$ into pieces $\Psi_k(x)$, evolve
each of these separately over time, and then recombine them to
construct the time-evolved $\Psi(x)$ itself. So far as the
\shro\ equation
is concerned, the division into $\Psi_k(x)$ pieces is arbitrary.
However, if the evolution of the $\Psi_k(x)$
is performed using \eqs{Rdotuni}{Sdotuni},
nonlinearity implies that the division is {\em not}
arbitrary, but has a large impact on the time-dependent
behavior of the resultant $R_k(x)$ and $S_k(x)$.

In principle, therefore, one may improve the numerical performance
of quantum trajectory calculations simply by judiciously dividing up
the initial wavepacket into several pieces.  It remains to be seen
the extent to which such a procedure will prove beneficial for
actual numerical calculations of real molecular systems---to be
sure, much depends on the manner in which $\Psi(x)$ is decomposed.
Nevertheless, the particular bipolar scheme explored in this paper,
already appears to exhibit much promise---at least with regard to
ameliorating the infamous node problem, which has thus far
severely limited the effectiveness of QTMs in the molecular arena.

The basic idea is to decompose a wavepacket, $\Psi(x)$, that
{\em has} nodes, into a linear combination of two components,
$\Ppm(x)$, that do not. In practice, it is not only nodes per se
that cause problems for numerical QTMs, but more generally,
any large or rapid oscillations in $\Psi(x)$.
Thus, $\Ppm(x)$ should ideally be not only nodeless, but also smooth
and slowly varying. For the special case where $\Psi(x)$
is a stationary Hamiltonian eigenstate, the semiclassical
method is well-known to yield approximate $\Ppm(x)$ functions with
the requisite properties---even when $\Psi(x)$ itself is highly
oscillatory. It is for
this reason that the semiclassical solutions were used as a guide
for determining the corresponding exact quantum $\Ppm(x)$'s.
Not only have we provided an explicit recipe for obtaining the
latter, we have also shown that these satisfy the
correspondence principle in the appropriate semiclassical limits.
Thus, the bipolar quantum potential, $q(x)$, obtained here---now understood
to represent the quantum correction to the semiclassical
approximation---is not only well-behaved in the vicinity of nodes,
but actually approaches zero, in the large action limit.

The bipolar quantum trajectories also behave very much like
classical/semiclassical trajectories; indeed, the two are
nearly identical in the classically allowed region, in the
large action limit. This may seem paradoxical, as both bipolar
and unipolar quantum trajectories conform to the same dynamical law,
and the latter are known to behave very non-classically. Unipolar
trajectories do not cross in position space, for instance---which can
cause kinky trajectories and other node-related difficulties,
especially when the wavepacket undergoes reflection. The
bipolar trajectories get around this difficulty as follows:
whereas the trajectories {\em on a single LM sheet} never cross
each other, they are all headed in the same direction anyway,
and so they don't get in each other's way. On the other hand,
trajectories on one LM sheet are free to cross those on the other
sheet---just like the corresponding semiclassical trajectories.
From a philosophical standpoint, one might thus regard the present
bipolar decomposition to be more compelling than the standard unipolar
approach---although curiously, this stance would require one to
abandon Bohm's original pilot wave interpretation.

The above discussion anticipates future application of the present
ideas to arbitrary time-evolving wavepackets; but it must be borne
in mind that thus far, only stationary wavepackets have been
considered. It is encouraging that the bipolar decomposition
scheme outlined here was found to be preserved over time for
stationary states (Sec.~\ref{QSHJEintro}). On the other hand,
for the more general time-evolving case, the LMs themselves will
change over time, and the bipolar decomposition scheme itself
need not be preserved. It is therefore possible that the
initially nodeless $\Ppm(x)$'s may develop nodes over the course
of time. This need not cause difficulties in practice however,
because at any desired time, one is free to redecompose $\Psi(x)$
into new $\Ppm(x)$'s that {\em are} nodeless.

One of the most appealing aspects of the excited harmonic
oscillator results of Sec.~\ref{HOexcited} is the fact that
the bipolar functions remain smooth and slowly varying for all
values of $n$. Indeed, the LMs for all $n$ values resemble each
other, apart from a change of scale. This is very advantageous from
a numerical perspective, as it suggests that very few trajectories
would be needed to accurately compute $r_n(x)$ derivatives, if
a completely numerical propagation scheme were adopted. More to
the point: the number of trajectories required should be essentially
{\em independent} of the number of nodes. For sufficiently
large $n$, it should even be possible to perform an accurate calculation
with {\em fewer than $n$ trajectories}---a prospect that would be virtually
unheard of in a unipolar context. Note that since the bipolar trajectories
themselves are also much smoother than the unipolar trajectories,
far fewer time steps should be required in the bipolar case.

We conclude with a brief discussion of the prospects for multidimensional
systems. At present, it is not entirely
clear how best to apply bipolar decomposition to an arbitrary, real,
multidimensional wavepacket, $\Psi(x_1,\ldots)$.  For regular systems,
semiclassical theory suggests an essentially direct-product decomposition,
via pairs of action-angle coordinates. Generally speaking, however, it is
difficult to find these coordinates---unless the wavepacket is initially
separable, which is very often the case in
molecular applications. On the other hand, the factor-of-two bifurcation
applies to each degree of freedom separately, resulting in $2^D$
components total, where $D$ is the number of DOFs. This is clearly
undesirable for large $D$. A more effective strategy may be to
bifurcate $\Psi(x)$ along the {\em reaction coordinate only}. These
and other ideas will be explored in future publications.

\section*{ACKNOWLEDGEMENTS}

This work was supported by awards from The Welch Foundation
(D-1523) and Research Corporation.
The author would like to acknowledge Robert E. Wyatt for many
stimulating discussions, and for introducing him to
the work of Floyd, and of Faraggi and Matone.
David J. Tannor is also acknowledged.
Jason McAfee is also acknowledged for his aid in converting this manuscript to an electronic format suitable for the arXiv preprint server.


\begin{thebibliography}{10}

\bibitem{bowman86}
J.~M. Bowman, J.~S. Bittman, and L.~B. Harding, J. Chem. Phys. {\bf 85},  911
  (1986).

\bibitem{bacic89}
Z. Ba\v{c}i\'{c} and J.~C. Light, Annu. Rev. Phys. Chem. {\bf 40},  469
  (1989).

\bibitem{bramley93}
M.~J. Bramley and T. {Carrington, Jr.}, J. Chem. Phys. {\bf 99},  8519  (1993).

\bibitem{poirier99qcII}
B. Poirier and J.~C. Light, J. Chem. Phys. {\bf 111},  4869  (1999).

\bibitem{poirier00gssI}
B. Poirier and J.~C. Light, J. Chem. Phys. {\bf 113},  211  (2000).

\bibitem{yu02a}
H.-G. Yu, J. Chem. Phys. {\bf 117},  2030  (2002).

\bibitem{yu02b}
H.-G. Yu, J. Chem. Phys. {\bf 117},  8190  (2002).

\bibitem{wangx03b}
X.-G. Wang and T. {Carrington, Jr.}, J. Chem. Phys {\bf 119},  101  (2003).

\bibitem{poirier03weylI}
B. Poirier, J. Theo. Comput. Chem. {\bf 2},  65  (2003).

\bibitem{poirier04weylII}
B. Poirier and A. Salam, J. Chem. Phys. {\bf 121},    (2004), (in press).

\bibitem{poirier04weylIII}
B. Poirier and A. Salam, J. Chem. Phys. {\bf 121},    (2004), (in press).

\bibitem{madelung26}
E. Madelung, Z. Phys. {\bf 40},  322  (1926).

\bibitem{vanvleck28}
J.~H. {van Vleck}, Proc. Natl. Acad. Sci. U.S.A. {\bf 14},  178  (1928).

\bibitem{keller60}
J.~B. Keller and S.~I. Rubinow, Ann. Phys. {\bf 9},  24  (1960).

\bibitem{maslov}
V.~P. Maslov, {\em Th{\'e}orie des Perturbations et M{\'e}thodes Asymptotiques}
  (Dunod, Paris, 1972).

\bibitem{arnold}
V.~I. Arnold, {\em Mathematical Methods of Classical Mechanics} (Springer, New
  York, 1978).

\bibitem{gutzwiller}
M. Gutzwiller, {\em Chaos in Classical and Quantum Mechanics} (Springer-Verlag,
  New York, 1990).

\bibitem{littlejohn92}
R.~G. Littlejohn, J. Stat. Phys. {\bf 68},  7  (1992).

\bibitem{brack}
M. Brack and R.~K. Bhaduri, {\em Semiclassical Physics} (Addison-Wesley,
  Reading, 1997).

\bibitem{lopreore99}
C.~L. Lopreore and R.~E. Wyatt, Phys. Rev. Lett. {\bf 82},  5190  (1999).

\bibitem{mayor99}
F.~S. Mayor, A. Askar, and H.~A. Rabitz, J. Chem. Phys. {\bf 111},  2423
  (1999).

\bibitem{wyatt99}
R.~E. Wyatt, Chem. Phys. Lett. {\bf 313},  189  (1999).

\bibitem{wyatt01b}
R.~E. Wyatt and E.~R. Bittner, J. Chem. Phys. {\bf 113},  8898  (2001).

\bibitem{wyatt}
R.~E. Wyatt, {\em Quantum Dynamics with Trajectories: Introduction to Quantum
  Hydrodynamics} (Springer, New York, 2005).

\bibitem{bohm52a}
D. Bohm, Phys. Rev. {\bf 85},  166  (1952).

\bibitem{bohm52b}
D. Bohm, Phys. Rev. {\bf 85},  180  (1952).

\bibitem{takabayasi54}
T. Takabayasi, Prog. Theor. Phys. {\bf 11},  341  (1954).

\bibitem{zhao03}
Y. Zhao and N. Makri, J. Chem. Phys. {\bf 119},  60  (2003).

\bibitem{garashchuk02}
S. Garashchuk and V.~A. Rassolov, Chem. Phys. Lett. {\bf 364},  562  (2002).

\bibitem{garashchuk03}
S. Garashchuk and V.~A. Rassolov, J. Chem. Phys. {\bf 118},  2482  (2003).

\bibitem{herman84}
M. Herman and E. Kluk, Chem. Phys. {\bf 91},  27  (1984).

\bibitem{kay94}
K.~G. Kay, J. Chem. Phys. {\bf 100},  4377  (1994).

\bibitem{miller01}
W.~H. Miller, J. Phys. Chem. A {\bf 105},  2942  (2001).

\bibitem{dey98}
B.~K. Dey, A. Askar, and H. Rabitz, J. Chem. Phys. {\bf 109},  8770  (1998).

\bibitem{burant00}
J.~C. Burant and J.~C. Tully, J. Chem. Phys. {\bf 112},  6097  (2000).

\bibitem{bittner02b}
E.~R. Bittner, J.~B. Maddox, and I. Burghardt, Int. J. Quantum Chem. {\bf 89},
  313  (2002).

\bibitem{shalashilin00}
D.~V. Shalashilin and M.~S. Child, J. Chem. Phys. {\bf 113},  10028  (2000).

\bibitem{burghardt01a}
I. Burghardt and L.~S. Cederbaum, J. Chem. Phys. {\bf 115},  10303  (2001).

\bibitem{burghardt01b}
I. Burghardt and L.~S. Cederbaum, J. Chem. Phys. {\bf 115},  10312  (2001).

\bibitem{donoso01}
A. Donoso and C.~C. Martens, Phys. Rev. Lett. {\bf 87},  223202  (2001).

\bibitem{wyatt01}
R.~E. Wyatt, C.~L. Lopreore, and G. Parlant, J. Chem. Phys. {\bf 114},  5113
  (2001).

\bibitem{donoso02}
A. Donoso and C.~C. Martens, J. Chem. Phys. {\bf 115},  6309  (2002).

\bibitem{bittner02a}
E.~R. Bittner, J. Chem. Phys. {\bf 115},  6309  (2002).

\bibitem{hughes04}
K.~H. Hughes and R.~E. Wyatt, J. Chem. Phys. {\bf 120},  4089  (2004).

\bibitem{trahan03}
C.~J. Trahan and R.~E. Wyatt, J. Chem. Phys. {\bf 118},  4784  (2003).

\bibitem{kendrick03}
B.~K. Kendrick, J. Chem. Phys. {\bf 119},  5805  (2003).

\bibitem{pauler04}
D.~K. Pauler and B.~K. Kendrick, J. Chem. Phys. {\bf 120},  603  (2004).

\bibitem{morette52}
C. Morette, Phys. Rev. {\bf 81},  848  (1952).

\bibitem{holland}
P.~R. Holland, {\em The Quantum Theory of Motion} (Cambridge University Press,
  Cambridge, 1993).

\bibitem{floyd94}
E.~R. Floyd, Physics Essays {\bf 7},  135  (1994).

\bibitem{brown02}
M.~R. Brown, arXiv:quant-ph/0102102  (2002).

\bibitem{goldstein}
H. Goldstein, {\em Classical Mechanics}, 2nd  ed. (Addison-Wesley, Reading, MA,
  1980).

\bibitem{milne30}
W.~E. Milne, Phys. Rev. {\bf 35},  863  (1930).

\bibitem{floyd86}
E.~R. Floyd, Phys. Rev. D {\bf 34},  3246  (1986).

\bibitem{bohr13}
N. Bohr, Phil. Mag. (Series 6) {\bf 26},  857  (1913).

\bibitem{wilson15}
W. Wilson, Phil. Mag. {\bf 29},  795  (1915).

\bibitem{sommerfeld16}
A. Sommerfeld, Ann. Phys. (Leipzig) {\bf 51},  1  (1916).

\bibitem{poirier00qcI}
B. Poirier, Found. Phys. {\bf 30},  1191  (2000).

\bibitem{messiah}
A. Messiah, {\em Quantum Mechanics} (Dover, New York, 1999), p. 232.

\bibitem{arfken}
G. Arfken, {\em Mathematical Methods for Physicists} (Academic Press, New York,
  1985).

\bibitem{floyd82}
E.~R. Floyd, Phys. Rev. D {\bf 25},  1547  (1982).

\bibitem{floyd00}
E.~R. Floyd, Found. Phys. Lett. {\bf 13},  235  (2000).

\bibitem{bouda01}
A. Bouda, Found. Phys. Lett. {\bf 14},  17  (2001).

\bibitem{faraggi99}
A.~E. Faraggi and M. Matone, Phys. Lett. B {\bf 455},  357  (1999).

\bibitem{faraggi00}
A.~E. Faraggi and M. Matone, Int. J. Mod. Phys. A {\bf 15},  1869  (2000).

\bibitem{langer37}
R.~E. Langer, Phys. Rev. {\bf 51},  669  (1937).

\bibitem{berry72}
M.~V. Berry and K.~V. Mount, Rep. Prog. Phys. {\bf 35},  315  (1972).

\bibitem{morehead95}
J.~J. Morehead, J. Math. Phys. {\bf 36},  5431  (1995).

\bibitem{poirier99lan}
B. Poirier, J. Math. Phys. {\bf 40},  6302  (1999).

\end{thebibliography}
\end{document}